\documentstyle[graphicx]{mn}

\newcommand{\df}{{\sc df}}
\newcommand{\rms}{{\sc rms}}
\newcommand{\kpc}{{\rm kpc}}
\newcommand{\mpc}{{\rm Mpc}}
\newcommand{\etal}{et.al.}
\newcommand{\gapprox}{\,\rlap{\lower 2.5pt            
 \hbox{$\sim$}}\raise 1.5pt\hbox{$>$}\,}
\newcommand{\lapprox}{\,\rlap{\lower 2.5pt            
 \hbox{$\sim$}}\raise 1.5pt\hbox{$<$}\,}

\newcommand{\figref}[1]{{Fig.~\ref{#1}}}
\newcommand{\eqref}[1]{{Equation~(\ref{#1})}}
\newcommand{\scmr}[1]{{\scriptsize \mbox{#1}}}
\newcommand{\comment}[1]{}
\newcommand{\kms}{\mbox{km s$^{-1}$}}
\newcommand{\msun}{M_\odot}
\newcommand{\dopic}[1]{{#1}}

\title{Dynamics of the Boxy Elliptical Galaxy NGC 1600}
\author[Michael Matthias and Ortwin Gerhard]
       { Michael Matthias and Ortwin Gerhard\\
        Astronomisches Institut der Universit\"at Basel, Venusstr. 7, 
        CH-4102 Binningen, Switzerland}
\date{submitted 1998 December 22}

\pagerange{\pageref{firstpage}--\pageref{lastpage}}

\begin{document}

\maketitle

\label{firstpage}

\begin{abstract}

We use three--integral models to infer the distribution function
(\df) of the boxy E3-E4 galaxy NGC 1600 from surface brightness
and line profile data on the minor and major axes.
We assume axisymmetry and that the mass-to-light ratio is constant in
the central $\sim 1 R_e$. Stars in the resulting gravitational
potential move mainly on regular orbits.  We use an approximate third
integral $K$ from perturbation theory, and write the \df\ as a sum of
basis functions in the three integrals $E, L_z$ and $K$. We then fit
the projected moments of these basis functions to the kinematic
observables and deprojected density, using a non-parametric algorithm.

The deduced dynamical structure is radially anisotropic, with
$\sigma_\theta/\sigma_r\approx\sigma_\phi/\sigma_r\approx 0.7$ on the
major axis. Both on the minor axis and near the centre the velocity
distribution is more isotropic; thus the model is flattened by
equatorial radial orbits. The kinematic data is fit without need for a
central black hole; the central mass determined previously from
ground-based data therefore overestimates the actual black hole
mass. The mass-to-light ratio of the stars is $M/L_V = 6h_{50}$.

The anisotropy structure of NGC 1600 with a radially anisotropic main
body and more nearly isotropic centre is similar to that found
recently in NGC 1399, NGC 2434, NGC 3379 and NGC 6703, suggesting that
this pattern may be common amongst massive elliptical galaxies.  We
discuss a possible merger origin of NGC 1600 in the light of these
results.

\end{abstract}

\begin{keywords}
Galaxies: kinematic and dynamics  --
Galaxies: elliptical --
Galaxies: individual: NGC 1600  --
Galaxies: formation  --
Line: profiles
\end{keywords}

\section{Introduction}

To understand the distribution of stellar orbits in elliptical
galaxies is a fundamental problem in stellar dynamics. Elliptical
galaxies are dynamically hot stellar systems, i.e., the velocity
dispersion of the stars is generally larger than their rotational
velocity.  In these three--dimensional systems the phase-space
distribution function (\df) of stars must depend on classical and
non-classical integrals of motion (Schwarzschild 1979), and may
involve stochastic orbit building blocks (Merritt \& Fridman 1996).

An important parameter for the dynamics of ellipticals is the central
density slope.  Parameterising the central
density as $\rho(r) \propto r^{-\gamma}$, there appear to be two
groups of galaxies (Gebhardt \etal\ 1996): ellipticals with weak cusps
($0\lapprox\gamma\lapprox 1.4$, peak at $0.8$) and with strong cusps
($\gamma > 1.4$, peak at $1.9$). The cusp properties turn out to be
related to other properties of ellipticals. Kormendy \& Bender (1996)
divide ellipticals into two groups:
\begin{itemize}
\vspace{-\baselineskip}
\item giant, cored ellipticals: non-rotating, anisotropic, boxy, moderately
triaxial, with cuspy cores,
\item lower luminosity power law ellipticals: rotating, nearly isotropic,
oblate--spheroidal, disky, strong cusps.
\end{itemize}
There is a range of luminosity where both types occur.
The natural question is whether these two groups have different formation
histories (Faber \etal\ 1997).

Elliptical galaxies are generally believed to have formed by some
variant of a merging process, as part of the hierarchical formation of
structure in the Universe. Depending on circumstances this could have
been a multiple merger between galaxies in a group, a merger between
two about equal spiral galaxies, or a merger between a dominant
galaxy and several minor companions. Numerical simulations of such
merging processes have been published, e.g., by Weil \& Hernquist (1996), 
Barnes \& Hernquist (1996), and Dubinski (1998), respectively. 

The shape and dynamical structure of the final remnant elliptical
galaxy depends sensitively on the influence of the dissipational
component during the collapse (Barnes \& Hernquist 1996). Even a small
fraction of the mass in gas is sufficient to drive the evolution
towards axisymmetry: in these calculations, including $10\%$ of the
mass of the disks in the form of gas changed a near-prolate final
remnant with axis ratios 10:5:4 to a near-oblate one with axis ratios
10:9:6.  The remnants of dissipationless mergers are also expected to
evolve slowly towards axisymmetry (Merritt \& Quinlan 1998), driven by
their central supermassive black holes: most spheroidal galaxies are
now believed to contain a central black hole with a fraction of
$\lapprox 0.5\%$ of the spheroid mass (Richstone \etal\ 1998, Magorrian
\etal\ 1998). In both cases the mechanism responsible appears to be the
destabilisation of the box orbits by the deep potential well, first
studied by Gerhard \& Binney (1985). The evolution caused by the black
hole proceeds through a sequence of quasi--equilibria by stochastic
diffusion (Merritt \& Fridman 1996).

These theoretical expectations are consistent with the
results of Franx, Illingworth \& de Zeeuw (1991), who used
observations of minor axis rotation to show that most ellipticals are
likely to be near-axisymmetric, with the majority of near-oblate shape
and a smaller fraction of near-prolate shape. However, the
distribution of apparent axis ratios of the giant cored ellipticals is
inconsistent with their being precisely axisymmetric (Tremblay \&
Merritt 1996). The majority of ellipticals without significant minor
axis rotation, including NGC 1600, are thus likely to be near-oblate
triaxial objects.

While quantitative information about the expected internal kinematics
and phase space structure of evolved merger remnants is still scarce,
it is clear that comparing this with the orbit distributions inferred
from observations will give important constraints on the processes
that shape ellipticals.  We have therefore started a project to
determine the stellar distribution functions (\df s) of flattened
elliptical galaxies from kinematic data.

An essential part of our technique is is the use of an approximate
third integral of motion. Based on the results discussed above we approximate
the mass distribution and potential as axisymmetric.  We calculate an
effective third integral of motion for the regular regions of
phase--space (after Gerhard \& Saha 1991), and then seek a
distribution function over three integrals which matches a given set
of photometric and kinematic data. The method used to determine the
\df\ is non-parametric and includes regularisation of the \df.  In
this paper we describe the technique and, as a first case, analyse the
surface brightness, velocity dispersion and line profile data for the
non-rotating E3-E4 galaxy NGC 1600.  It is known that some ellipticals,
especially NGC~1600, can not be fitted by two-integral models
(Binney, Davies \& Illingworth 1990, van der Marel 1991).

The paper is organised as follows.  In Section 2 we briefly summarise
the observational data, and discuss our assumptions in Section 3. Our
technique to infer the \df\ is explained in detail in Section 4. The
results for NGC 1600 and a discussion follow in Sections 5 and 6,
respectively.

\section[]{Observational Data}

NGC~1600 is a bright ($M_B=-23.17$) elliptical galaxy at a distance of
$D=93\mpc$ (for $H_0=50\kms/\mpc$).  To derive the stellar density
distribution we have used surface photometry of NGC~1600 from Bender
(private communication). The effective radius is $R_e=48"$ ($
21.6/h_{50}\kpc)$.  To constrain the dynamical models we have used
velocity dispersions and line profile shape parameters measured by
Bender, Saglia \& Gerhard (1994). Older velocity dispersion data by
Jedrzejewski \& Schechter (1989) do not contain line profile shape
information and show systematically lower velocity dispersions in the
region of overlap, especially in the center. We have therefore decided
to use only the newer data by Bender \etal\ (1994) in the modelling.
These data extend to approximate $R_e/2$ along both the major and
minor axes; the data are binned to eight points on the major axis and
eleven on the minor axis.  The line profile shapes are expanded in
Gauss-Hermite moments.

NGC~1600 shows only little rotation. The maximal rotation velocity is
around $30 \kms$ but most of the measured rotational velocities are
below $20\kms$ with errors of comparable size. Consistent with the
lack of rotation the skewness of the line profiles is nearly zero,
except at one radius where $h_3 \approx -0.15$. In the following we
therefore use only the $\sigma$ and $h_4$ data.

\section{Assumptions}

As discussed in the Introduction, both theory and observation suggest
that old giant ellipticals are triaxial but not far from axisymmetric.
The small measured rotation velocities and $h_3$ parameters on the
minor axis then suggest that NGC 1600 is near--oblate.  Therefore we
assume an oblate--axisymmetric mass distribution and potential. This
assumption is undoubtedly only an approximation -- the results of
Hunter \& de Zeeuw (1992) indicate that we should expect about $20\%$
of the total mass of stars on box-like and $x$-tube orbits if NGC 1600
is triaxial with axis ratios $\sim10:9:6$ -- but it makes the
subsequent analysis much easier and is likely to give approximately
correct kinematic results.  We will see whether the data for NGC 1600
can be fit by an axisymmetric model or whether triaxiality is required
by the kinematics.

The projected axis ratio of NGC~1600 is E3-E4. Intrinsically flatter
cored ellipticals are rare (Tremblay \& Merritt 1996), thus NGC 1600
is likely to be nearly edge--on.  We therefore assume an inclination
angle of exactly $i=90^\circ$; this ensures that the deprojection of
the surface density is unique. If the geometry is not edge--on,
disk--like konus densities can be added to the density without
altering the surface brightness (Gerhard \& Binney 1996); however, the
resulting uncertainty in the three-dimensional density distribution
decreases to zero as $i\rightarrow 90^\circ$.

The small measured rotation velocities and $h_3$ parameters on the
major axis of NGC~1600 are consistent with the assumption that this
galaxy is nonrotating. In this case, the skewness (and all higher odd
Gauss-Hermite moments) of the line profiles vanish and we remain with
the velocity dispersion and $h_4$ kinematical data. Non-rotating
models have \df s even in $L_z$, so that we restrict ourselves to
models even in $L_z$.

Finally, we assume a constant (but free) mass-to-light ratio $M/L$ in
the central $ R_e/2$ where we have kinematical data. I.e., we assume
that in this region the high density of stars dominates over the dark
matter density.  This assumption appears reasonable in view of the low
central $M/L_B=3.3$ inferred for the E0 galaxy NGC~6703 even when the
stars have the maximum mass compatible with the central kinematics
(Gerhard \etal\ 1998). In this galaxy the dark halo does not become
important until $\approx 1 R_e$.

\section{The Method}

Our aim is to obtain a three-integral distribution function for
NGC~1600 by modelling all data (surface brightness and line profiles
on two axes) simultaneously in a $\chi^2$-sense.  Because the third
integral is calculated from the perturbation away from a spherical
{\em potential}, we derive in a first independent step the
three-dimensional density distribution from the observed surface
brightness and then the gravitational potential from the
density. Under the assumptions made we obtain a unique $\rho(R,z)$ and
$\Phi(R,z)$.  Given the potential, the approximate third integral is a
series expansion in the action-angle variables of the spherical part
of $\Phi(R,z)$ (see Gerhard \& Saha 1991).

We then set up basis functions for the \df\ that depend on the three
integrals $E$, $L_z$, and the third integral $K$.  For each basis
function we calculate moments which are projected along the
line-of-sight. The \df\ is written as a sum over these basis functions
and its projected moments are linear sums over the corresponding
moments for the basis functions. The coefficients are determined by
fitting directly to the observations, except for the model densities
which are fitted to the already deprojected $\rho(R,z)$.

The determination of the \df\ from observations by this process is an
ill-posed problem (see, e.g., Merritt 1993), because the observables
are integral moments of the underlying \df, and so small changes in
the observables lead to large changes in the recovered \df\
(spikes). Typically, the inferred function becomes spiky because of
the amplification of structures induced by measurement errors. To
avoid this we include a regularisation term in the $\chi^2$--function;
this is minimized by a linear fitting routine within the usual
non-negativity constraints and mass conservation.  In this way a
smooth \df\ consistent with the data is found.

Our approach of analyzing line profile data for axisymmetric galaxies
is related to that of van der Marel \etal\ (1998), Gebhardt \etal\
(1999), and Cretton \etal\ (1999) who use a generalized version of
Schwarzschild's method, and to that of Dejonghe \etal\ (1996) and
Emsellem \etal\ (1999) who use a St\"ackel integral obtained from
fitting a single St\"ackel potential to the potential of the galaxy
under study.  The difference is that we use a third integral derived
from the galaxy potential rather than following individual orbits, and
that we take care to investigate how well our third integral actually
represents the orbits in this potential. In the following we describe
the different steps in our method and their application to NGC 1600 in
more detail.

\subsection{Deprojection and gravitational potential}
\label{depdens}
With the above assumptions the deprojection is unique and the
gravitational potential is determined by the density up to a constant
factor.  We used a program by Dehnen (1995), which calculates the
density $\rho(R,z)$ from the surface brightness $\Sigma(X,Y)$ by a
Lucy-algorithm (Lucy 1974), and then evaluates the potential
$\Phi(R,z)$ for constant $M/L$ ratio as a sum of spherical
harmonics. Here $(X,Y)$ are sky-coordinates and $(R,z)$ are
coordinates in the meridional plane of the galaxy. Between Lucy steps
the density is smoothed using FFT filtering.

The calculation is done on a grid where the grid points lie on $11$
rays through the ($R\ge0$, $z\ge0$) quadrant of the meridional plane,
including the two axes. The grid extends to a maximum radius of $30
\kpc$. For NGC 1600, the density is thus extrapolated slightly beyond
the edge of the CCD data (corresponding to a galactocentric radius of
$28\kpc$). The extrapolation assumes a power-law with exponent
$\gamma=-4$ at large radii.

Fig.~1 shows density profiles along three axes resulting from this
deprojection.  The deprojected central density slope is $\gamma=0.24$,
consistent with the result found by Gebhardt \etal\ (1996). NGC 1600 is
the only galaxy in their sample which is consistent with a flat
central density profile. The mean axial ratio of the deprojected
density distribution in Fig.~1 is $c/a=0.68$.

\begin{figure}
\dopic{\includegraphics[width=0.45\textwidth]{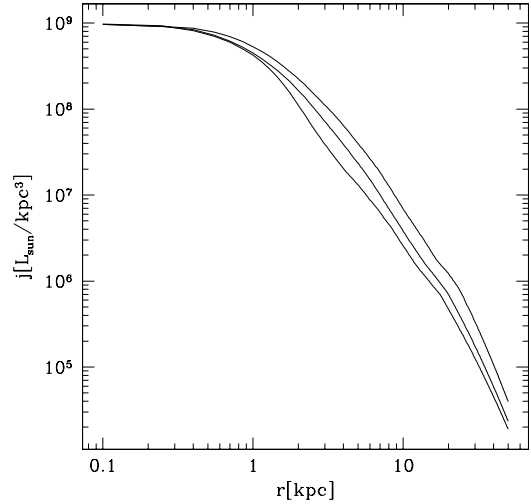}}
\caption{
\label{denspic}
Three-dimensional luminosity density of NGC 1600 along the minor, major and
an intermediate axis, from deprojecting the observed surface brightness
distribution with $i=90^\circ$.
}
\end{figure}

\begin{figure}
\dopic{\includegraphics[width=0.45\textwidth]{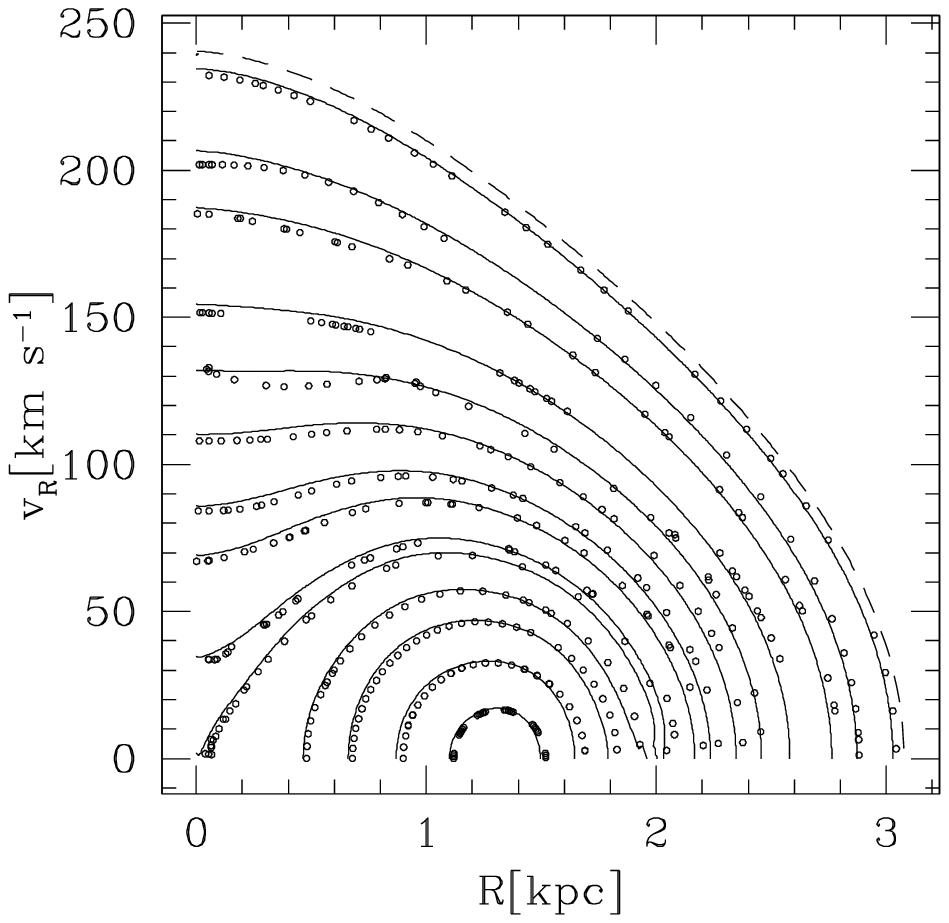}}
\dopic{\includegraphics[width=0.45\textwidth]{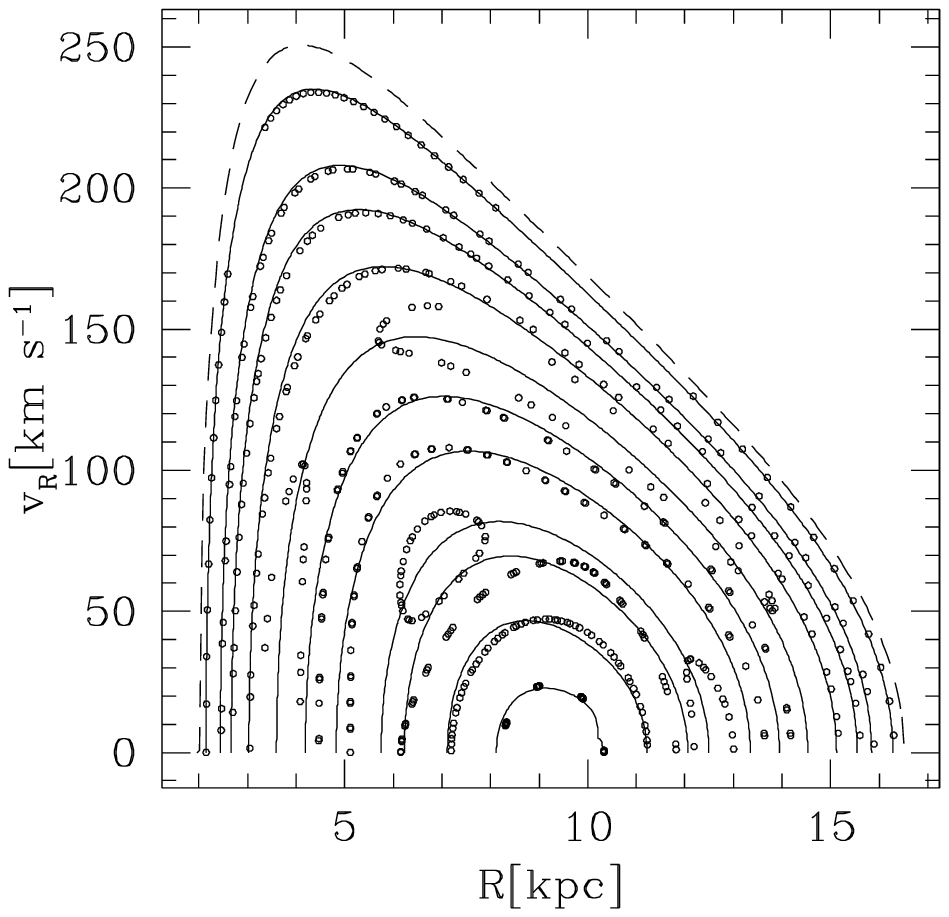}}
\caption{
\label{sospic}
Top: A typical surface of section for $L_z=0$ orbits in NGC~1600. 
The squares represent numerically integrated orbits, whereas 
the overlayed lines are contours of constant third integral $K$.
The energy for this SOS was chosen so that the radius of the
circular orbit in the equatorial plane is $R_c=2\kpc$. Bottom:
high-energy SOS with $L_z/L_{\rm circ}(E)=0.4$; here $R_c=10\kpc$. 
}
\end{figure}

\subsection{Third integral}
\begin{figure*}
\dopic{\center{\includegraphics*[width=.8\textwidth]{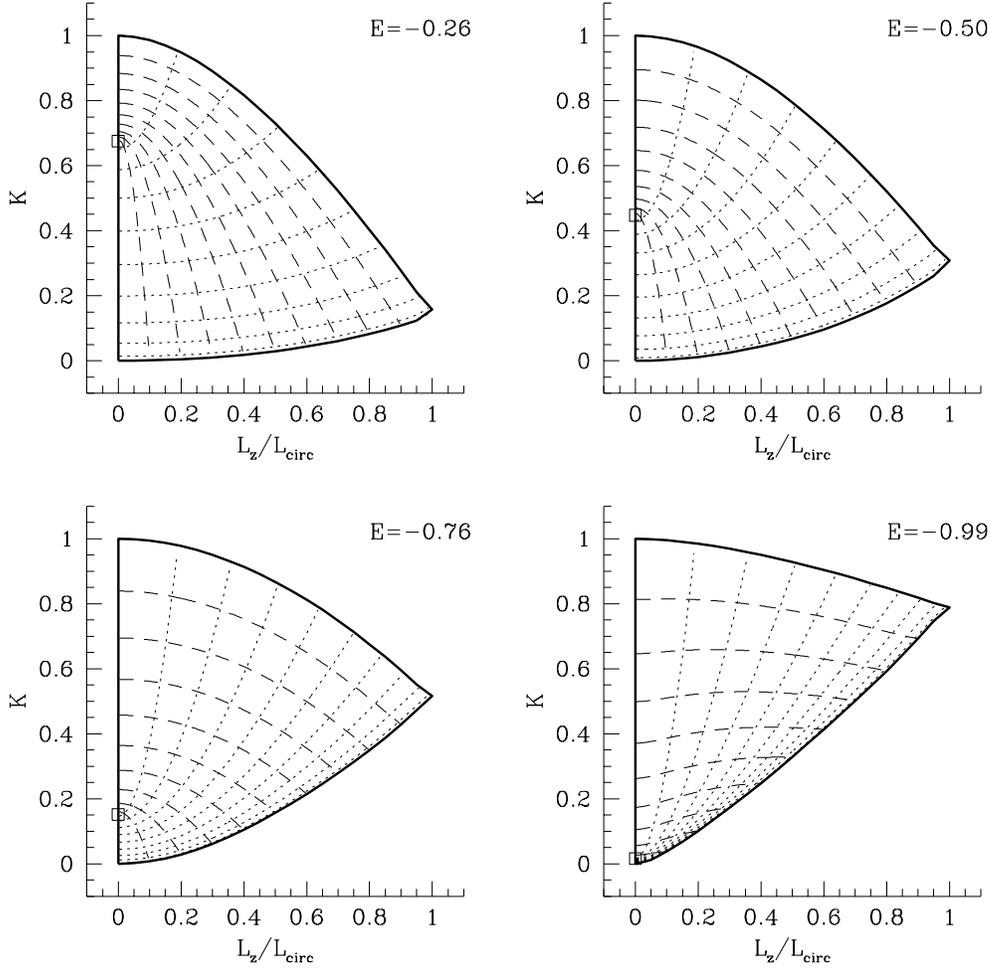}}}
\caption{\label{isoldpic}
Integral space for four different energies ranging from $E=-0.26$ (top left)
to $E=-0.99$ (bottom right). The circular orbit radii corresponding to
these energies are $R_c=16\kpc, 6\kpc, 2.2\kpc, 0.5\kpc$.
In these diagrams the circular orbit in the equatorial plane is at the
right hand corner, the equatorial radial orbit at the top left, and the
closed meridional loop orbit at the lower left corner. Equatorial orbits
lie on the upper right boundary, shell orbits on the lower right boundary,
meridional butterfly orbits on the upper part and meridional loop orbits
on the lower part of the left hand boundary.
The box on the $L_z=0$ axis denotes the critical orbit at $K_\scmr{crit}$
dividing the latter two families.
The dotted lines represent the shape invariant $S_r$, the dashed lines
$S_m$. Note the crowding of contour lines near $K_\scmr{crit}$.
}
\end{figure*}

In an axisymmetric potential the classical integrals of motion are the
energy $E$ and the $z$-component of the angular momentum $L_z$.
Numerical orbit integrations show, however, that particles in the
potential of NGC 1600 obey an approximate third integral of motion
(see below), which we call $K$.  Gerhard \& Saha (1991) developed a
method to calculate an approximation for the third integral.  This is
based on resonant perturbation theory and uses a Lie-transform of the
unperturbed integrals in terms of the action--angle variables of the
unperturbed spherical part of the potential. The expression obtained
for the third integral usually is a good approximation if the density
is rounder than $c/a\simeq 0.5$.  Dynamical models making use of this
third integral were studied by Dehnen \& Gerhard (1993a,b) for a
perturbed isochrone sphere. 

Here the algorithm has been generalised for the potentials of real
galaxies. The deprojected density distribution of the galaxy is
expanded in spherical harmonics and the corresponding gravitational
potential is derived.  As unperturbed integrable potential we take the
spherical part of this potential expansion, and for the perturbation
we here take the $l\!=\!2$ and $l\!=\!4$--terms, but excluding a 4:3
resonance term. The resulting resonant invariant describes the 2:1
resonant (in terms of the frequencies of the spherical system)
meridional butterfly orbits, and is of sufficient accuracy to describe
most z--tube orbits well. We make no attempt to
treat the 4:3 and other resonant orbit families specially; by using
our resonant invariant we effectively fit a tube orbit torus through
each chain of resonant islands.

Fig.~\ref{sospic} illustrates this by showing two typical surfaces of
section for the potential of NGC 1600, calculated by numerically
solving the equations of motion.  Overlayed are the contours of $K$
calculated by perturbation theory, at values of $K$ corresponding to
the mean values of $K$ along each orbit. The top panel shows
zero--angular momentum orbits at an energy corresponding to a circular
orbit radius of $R_c=2\kpc$.  The agreement is excellent; the error of
the third integral along an orbit is typically two percent.  A
corresponding $L_z=0$ surface of section at $R_c=10\kpc$ looks
similar: there are still almost no stochastic orbits and the
description of the invariant curves by the third integral $K$ is
similarly good.  The bottom panel of Fig.~\ref{sospic} shows
higher--angular momentum orbits ($L/L_{\rm circ}(E)=0.4$) at an energy
corresponding to $R_c=10\kpc$.  There are now some significant
families of resonant islands which are not fit by the resonant
invariant as used here, but are approximated by tube orbit tori.

For a steady--state galaxy the strong Jeans theorem states that the
distribution function $f$ of the stars depends on the three
independent integrals of motion only, provided all stars move on
regular orbits with incommensurable frequencies (Binney \& Tremaine
1987).  For NGC~1600 this is nearly the case as Fig.~\ref{sospic}
shows. Therefore we now seek a function $f(E,L_z,K)$ which can
reproduce all available observations for NGC~1600, i.e. surface
density, velocity dispersion and line profile shapes.

\subsection{Integral Space}
\label{secISpace}
The classical integrals $E$ and $L_z$ together with the third integral
$K$ form a complete set of coordinates for the integral space (Dehnen
\& Gerhard 1993a). In integrable potentials, where the third integral
is conserved exactly, each point of the integral space represents a
single orbital torus.  In the present case where the third integral is
only an approximation, albeit a good one (Fig.~\ref{sospic}), this is
nearly true.

For fixed energy, the integral space has a triangular shape, defined by the
range of values taken by the two other invariants $L_z$ and $K$. In the
representation shown in \figref{isoldpic}, $L_z$ is normalised by the
angular momentum of the circular orbit in the equatorial plane.  In the
plots in \figref{isoldpic}, the equatorial circular orbit is thus located at
$L_z/L_\scmr{circ}=1$ in the right hand corner.  The adjacent boundary at
high values of $K$ delineates equatorial orbits with radial action increasing
to the left, the boundary at low values of $K$ represents shell orbits with
vertical excursions increasing towards the lower left. Orbits with $L_z=0$ can
be separated into two groups, depending on their values of $K$. Those with
$K$ less than a critical value $K_\scmr{crit}$, indicated by the square
on the left boundary, are meridional loop orbits, those above the critical
value are meridional butterflies that shrink vertically as the
equatorial radial orbit is approached in the upper left hand corner. 

\begin{figure}
\dopic{\center{\includegraphics*[width=.45\textwidth]{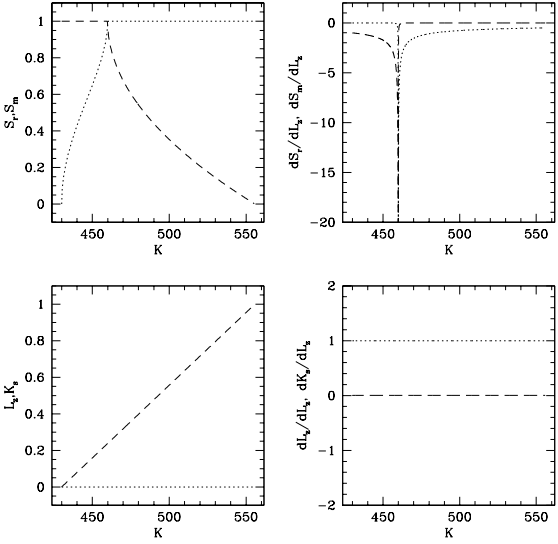}}}
\caption{\label{dsmsrpic}
Old (top panels) and new invariants (bottom panels)  on
the $L_z=0$ axis for some fixed energy.
The left column shows the invariants and the right column the derivatives
with respect to $L_z$ and $K$. Clearly visible is the singular behaviour of
the old invariants at $K=K_\scmr{crit}$,  which the new invariants
$L_z$ and $K_s$ do not show.
}
\end{figure}

Using the value of $K_\scmr{crit}$, shape invariants $S_r,S_m$ can be
constructed from $L_z, K$ such that they describe the radial and
meridional extent of the orbits (these are approximate turning point
variables; see Dehnen \& Gerhard 1993a,b). The critical value
$K_\scmr{crit}$ depends on energy; below a certain energy
$K_\scmr{crit}$ is identical with $K_\scmr{min}(0)$, the minimum value
of $K$ at $L_z=0$.  In this case the contours of $S_m$ cluster at
$K_\scmr{min}(0)$ and the area covered by each 'box' between contour
lines in Fig.~\ref{isoldpic} vanishes.  In addition, the shape
invariants become singular at the critical $K_\scmr{crit}$, i.e.,
their derivatives with respect to $L_z$ and $K$ are indefinite at this
point (Fig.~\ref{dsmsrpic}).  Because we need a smooth representation
of the integral space and a smooth and differentiable phase space
distribution function $f$, we have therefore constructed a new
representation of integral space.

\begin{figure*}
\dopic{\center{\includegraphics*[width=.8\textwidth]{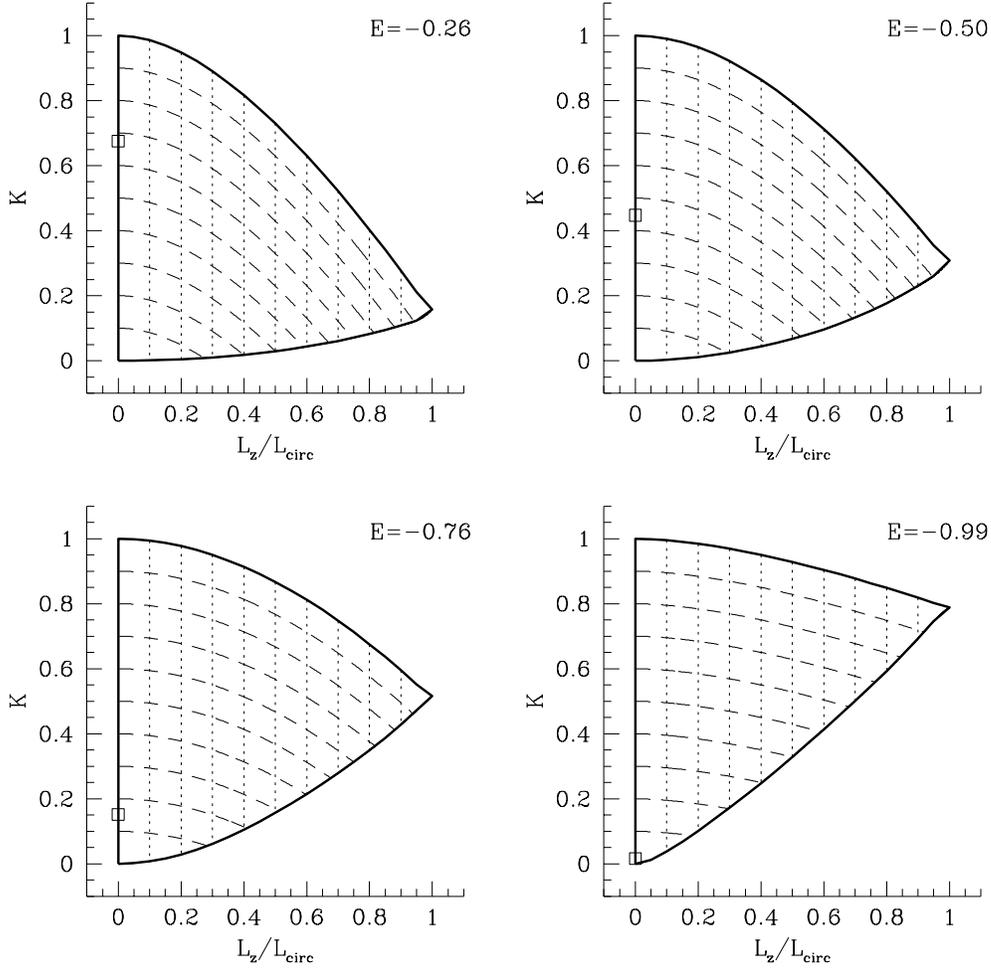}}}
\caption{\label{isnewpic}
Same as Fig.~\ref{isoldpic} but for the new invariants $L_z$ and $K_s$;
see text. Now the integral space is divided by the contours into regular cells
and the crowding of contours near the critical orbit has disappeared.
}
\end{figure*}

To this end we introduce a new quantity
\begin{equation}
K_s = 1 - \frac {K_\scmr{max}(L_z)-K}
                {K_\scmr{max}(0)-K_\scmr{min}(0)}.
\end{equation}
Contours of $K_s$ are contours of scaled $K_\scmr{max}(L_z)-K$ : the upper
boundary is shifted downwards according to the value of $K$.
Fig.~\ref{isnewpic} shows contours of the new set of invariants $L_z$ and
$K_s$ on several energy surfaces through integral space.
On the upper boundary line in these diagrams equatorial orbits have
$K_s=1$, and for the closed meridional loop in the lower left hand corner of
integral space $K_s=0$. Now the area
covered by each box between contour lines is approximately constant and no
singularities appear.
In the following we use the invariants $E$, $L_z$ and
$K_s$ as a representation of integral space.

\subsection{Distribution Function}

The \df\ is written as a sum over basis functions $f_l(E,L_z,K_s)$:
\begin{equation}
  f(E,L_z,K_s) = \sum_{l=1}^{l_{\rm max}} \alpha_l f_l(E,L_z,K_s).
 \label{FAsASum}
\end{equation}
Suitable basis functions are constructed using the separation ansatz
\begin{equation}
f_l(E,L_z,K_s) = f_{ij}(E,L_z,K_s) = g_i(E) \times h_j(L_z,K_s).
\end{equation}
The functions $g_i(E)$ describing the energy-dependence of the \df\ are
determined as follows. First, we construct an isotropic \df\
$f_\scmr{iso}(E)$ whose zeroth moment approximates the spherically
averaged density profile of the galaxy. This function is used as the
basic energy function; further energy functions are constructed by
multiplying this isotropic function by binomials centred at different
energies $E_i$. The $E_i$ are chosen such that the corresponding
$g_i(E)$ probe different regions in energy and cover the total energy
range approximately uniformly. In this paper, we use seven energy
basis functions $g_i(E)$.

The basis functions $h_j(L_z,K_s)$ that describe the orbit
distribution on energy surfaces are constructed using powers of the
(new) invariants $|L_z|^m K_s^n$ with $n+m \le 4$. These 15 angular
basis functions plus two additional isotropic components $\propto E,
\propto E^2$, are multiplied by all of the energy basis functions,
giving a set of $l_{\rm max} = 7\times (15+2) = 119$ basis functions
$f_{l}(E,L_z,K_s)$.  The \df\ is a linear combination of these $f_l$
with weights $\alpha_l$; cf.\ ~\eqref{FAsASum}.

\subsection{Velocity space and line-of-sight integration}

All observables are line-of-sight projections of the intrinsic
quantities. E.g., the surface brightness $\Sigma(X,Y)$ at a
point $(X,Y)$ on the sky is given by $ \Sigma(X,Y) =
\int_{-\infty}^{\infty} dZ \rho(X,Y,Z)$, where $Z$ is the coordinate
along the line of sight and $\rho(X,Y,Z)$ the intrinsic density at
$(X,Y,Z)$. This itself is an integral over velocity
space: $\rho(X,Y,Z) = \int_{-\infty}^{\infty} d^3v f(\vec r, \vec v)$,
with $\vec r = (X,Y,Z)$ and $f(E,L_z,K_s)$ the distribution function.
Thus we may write $\Sigma = {\bf S} \, f$, where the operator ${\bf S}
\equiv \int_{-\infty}^{\infty} dZ d^3v $. The same operator is needed to
calculate the projected velocity dispersion and Gauss--Hermite moments
from the \df. Here we describe the evaluation of the velocity and
line--of--sight integrals.

We perform the integrations over velocity space in the manner of
Dehnen \& Gerhard (1993a). Using axisymmetry, and after a
transformation of the integration variables $(d^3v \rightarrow dE dL
dL_z)$, the operator ${\bf R}\equiv \int_{-\infty}^\infty d^3v$ becomes
\begin{eqnarray}
{\bf R} \equiv \int_{-\infty}^{\infty}\left .  d^3v\,\right|_{R,z}
  &=&          \frac 1r \int_{\Phi(R,z)}^{E_\scmr{max}}
               dE \int_0^{r\sqrt{2[E-\Phi(R,z)]}} \nonumber \\
  &\times&     \frac{L dL}{\sqrt{2[E-\Phi(R,z)] r^2 - L^2}} \nonumber \\
  &\times&     \int_{-\frac {R}r L}^{\frac Rr L} 
               \frac{dL_z}{\sqrt{\left( \frac Rr L \right)^2 - L_z^2}}
               \sum_{{sgn(\dot r)=\pm} \atop {sgn(\dot \theta)=\pm}}.
\label{defineR}
\end{eqnarray}
As usual, $R$ and $z$ are cylindrical coordinates, $r^2=R^2+z^2$, and
$\Phi(R,z)$ is the gravitational potential in the meridional plane.
The maximal energy $E_\scmr{max}$ appearing as the upper integration
boundary should be the value of the gravitational potential at
infinity, but in practice is the value of the potential at a distance
of $30\kpc$ along the major axis (see \S\ref{depdens}).

The operator ${\bf R}$ is applied to functions of velocity (depending
on the desired moment) times the basis functions $f_{ij}(E,L_z,K_s)$
which involve the third integral $K$.  Because the computation of $K$
is time consuming, values of $K$ are pre-calculated on a grid in $E$,
$L$ and $L_z$ for each point of the grid in the meridional plane.
Thus the integrand can be evaluated only on the grid points, and we
have therefore performed the integrations over $L_z$ and $L$ by
Gauss-Tschebyschev- and Gauss-quadrature, respectively. For the
remaining integral over energy we use spline-interpolation and a
Bulirsch-Stoer integrator.

The velocity integration yields the intrinsic moments like the density
and velocity dispersions in the meridional plane of the galaxy. These
are integrated along the line of sight to obtain the observable
moments.  For this we also use a Bulirsch-Stoer integrator and
interpolate bilinearly in the meridional plane for the minor axis
kinematic data, and in the equatorial plane for the major axis data.

\subsection{The fitted velocity moments}

Having obtained the intrinsic and projected moments of all basis
functions, the \df\ for a galaxy can now be found by matching to the
observed moments. The quantities included in the fit are the density,
the velocity dispersions on the major and minor axes, and the measured
line profile parameters ($h_4$ in the case of NGC 1600, and possibly
$h_3$ and higher $h_n$). Some details are described in this
Subsection.

{\bf density:} The model is required to fit the deprojected
three-dimensional brightness distribution
\begin{equation}
\rho(R_i,z_i) = \sum_l \rho_l(R_i,z_i)
\label{DensityAsASum}
\end{equation}
 on a grid $(R_i,z_i)$ in the meridional plane, with $\rho_l(R_i,z_i) =
{\bf R}\, f_l(E,L_z,K_s)$.  The employed grid is similar to that used
in the deprojection in that the grid points lie on $11$ rays through
the ($R\ge0$, $z\ge0$) quadrant of the meridional plane, including the
two axes. This grid extends to a maximum radius of $26.6 \kpc$.  The
large range in radius allows us to estimate the contribution of
high--energy stars on near--radial orbits to the kinematic moments
further in.

{\bf velocity dispersion:} The velocity dispersion given by Bender
\etal\ (1994) is {\sl not} the second moment of the \df\ $\sigma_0$ but
corresponds to the parameter $\sigma_\scmr{fit}$ in a several
parameter {\sl fit} function for the entire line profile, including
the $h_3$ and $h_4$ terms. Only when $h_3=h_4=0$ will $\sigma_0$ be
equal to $\sigma_\scmr{fit}$.  Thus we first determine the second
moments of the observed line profiles by integrating over the line
profile $l(v_{||})$ as specified by $(\sigma_\scmr{fit}, v=0, h_3=0,
h_4)$. For negative $h_4$, we integrate only up to the velocity where
$l(v_{||})$ first becomes negative. Setting $v=h_3=0$ assumes that
rotation is negligible.

Given estimates for $\sigma_0$, the model is required to satisfy
\begin{equation}
 \Sigma \sigma_0^2 = \sum_l \alpha_l \Sigma_l \sigma_{ll}^2
\label{SigmaAsASum}
\end{equation}
at the positions of all data points. Here $\sigma_{ll}$ is the
projected velocity dispersion of the basis function $l$, $\Sigma_l$
its surface brightness, and the total surface brightness
$\Sigma = {\sum_l\alpha_l \Sigma_l }$ from \eqref{DensityAsASum}.

{\bf Gauss-Hermite parameter $h_4$:}
The measured line profile parameters $h_4$ depend non-linearly on the
galaxy's \df\ and can therefore not directly be used in a linear
least squares algorithm. We therefore transform to a new set of
even Gauss-Hermite moments $s_n^{(\hat v, \hat \sigma)}$,
using fixed fiducial velocity scales
$\hat v(X_i,Y_i)$ and $\hat \sigma(X_i,Y_i)$, where $(X_i,Y_i)$
denote the position of the $i^{\rm th}$ data point on the sky plane.
These fiducial velocities are taken from a dynamical model that
approximately matches the observed velocity dispersions
and has $\hat v(X_i,Y_i)=0$. This ensures that
the transformed Gauss-Hermite series converge quickly. 

Expressed in terms of the new s-moments,
the observed line profile shapes now depend linearly on
the \df:
\begin{equation}
 \Sigma s_n^{(\hat v, \hat \sigma)}=
	\sum_l \alpha_l \Sigma_l s_{n,l}^{(\hat v, \hat \sigma)}
\label{SnAsASum}
\end{equation}
with $s_{n,l}^{(\hat v, \hat \sigma)}$ the Gauss-Hermite moments of the
$l^{\rm th}$ basis function evaluated with the same velocity scales
$(\hat v, \hat \sigma)$. We have calculated and required the model to
match s-moments up to order $s_6$.

\subsection{Linear $\chi^2$ fitting including regularisation }

The constraint equations (\ref{DensityAsASum} -- \ref{SnAsASum})
involve the integral operators ${\bf R}$ and ${\bf S}$. Inferring the
\df\ by solving these equations is an ill-posed problem in the sense
that small changes in the observational data can lead to large
variations in the inferred function. To prevent artificial spiky
structures in the inferred \df\ generated by noise in the data one has
two principal possibilities (e.g., Scott 1992, Merritt \& Tremblay
1994).  One is to try a {\sl parametric} inversion, i.e., fitting a
function $f_p(r | a,b,c,...)$ with a small number of parameters
$a,b,c,...$. The resulting $f_p$ is always smooth, but because the
fixed functional form of $f_p$ may not be suitable for the true
function $f(r)$, features may be induced which are not real (bias).

The other is to use a {\sl non-parametric} inversion, where the
inferred function is represented by a large number of elements
$\alpha_i$ (values on grid points, basis functions, etc.) that give
$f_{np}(r|\alpha_i)$ the freedom to match any function $f$.  Such an
inversion must be regularised, for otherwise the fit to the data will
be too good ($\chi^2$ per point $\ll 1$) and the resulting
$f_{np}(r|\alpha_i)$ will contain unphysical structure that depends
purely on the noise in the data.  A common method is to restrict the
curvature or second derivative of the inferred function
(Wahba \& Wendelberger 1980). Instead of
the usual $\chi^2$--function, one minimizes the quantity
\begin{eqnarray}
 \xi^2 &=& \chi^2 + \lambda P(f)\nonumber \\
       &=& \frac 1n \sum_j \frac{[O_j-O(r_j)]^2}{\sigma_j^2}
        + \lambda \int_{0}^\infty dr \, [f''(r|\alpha_i)]^2,
\end{eqnarray}
where $O_j$ denotes the $j^{th}$ measurement at position $r_j$ with
error $\sigma_j$. $O(r_j)$ represents a linear operator which relates
the function space to the observable space; in our case this will be
${\bf R}$ and ${\bf S}$. The fitting of the data is done in the space
of the observations, whereas the regularisation happens in the
intrinsic space of the \df. The parameter $\lambda$ determines the
amount of regularisation: for $\lambda=0$ the standard $\chi^2$
fitting is recovered, for $\lambda=\infty$ the result is determined by
the regularisation function. In the case given one obtains a linear
function whose slope and offset are determined by the data.  In
astronomical applications, the data often do not sample the desired
functions very well. Then it is necessary to use relatively large
values of $\lambda$, and so the result will again be somewhat biassed,
by the form of the regularisation term.

We have chosen the second approach as the one that will adapt more
easily to future large and accurate kinematic data sets. We use basis
functions rather than grid cells in integral space because, due to the
complexity of the third integral $K_s$, the complicated phase-space
boundaries of such grid cells make it difficult to apply the
operators ${\bf R}$ and ${\bf S}$. The number of basis functions is
adapted to the NGC 1600 data, but this is easy to change.

Because all projected moments depend linearly on the \df, we use the
constrained linear least squares {\it netlib} routine {\sl lfit}
(Hanson \& Haskell 1981), which
solves, in a $\chi^2$ sense, a set of equations $\vec y = {\bf A} \vec
x + \vec b $, subject to linear equality ($\vec y = {\bf E} \vec x +
\vec c$) and inequality constraints $\vec y \leq {\bf U} \vec x + \vec
d$. In our case the matrix ${\bf A}$ consists of equations for the
density, the velocity dispersion, the first three even $s$-moments and
the regularisaton terms. The latter can be included in the linear
fitting routine because the penalty function $P(f)$ has a quadratic
form; in practise we ask the routine to solve $f''=0$ on a grid in
integral space, again in a $\chi^2$ sense and suitably weighted by
`errors' $\lambda^{-1/2}$.  For the employed grid this gives rise to
$5184$ additional linear equations.

The only equality constraint we have included in the matrix ${\bf E}$
is a luminosity (or mass) conservation constraint. The need for this
arises because of the smoothing term and because the density is fitted
only in a $\chi^2$-sense. For fixed $\lambda$, the model's penalty
function $P(f)$ can be reduced by either decreasing the curvature of
the model or by multiplying all basis functions $f_l(E,S_r,S_m)$ by a
fixed number less than one.  Depending on the shape of the $\xi^2$
hyper-surface, it is possible that scaling of the model is favoured
over reducing the curvature.  To ensure that the total brightness of
the model remains equal to that of the galaxy, we add an equality
constraint which forces the solution to have the same total luminosity
as the observed galaxy.

Finally, the non-negativity of the \df\ is imposed on a grid
of $10920$ points in
$E,K$ and $L_z$ and defines the components of matrix ${\bf U}$.

The weights for the several fitted quantities are determined as
follows: For the relative weights of the dispersion and s-moments we
use the values of Gerhard \etal\ (1998) determined by Monte Carlo simulations.
Minor and major axis kinematic data have the same weights. The
relative weights of the density and kinematics were chosen in such a
way that the overall RMS error in the density was less than $1\%$. The
final free parameter is the smoothing parameter $\lambda$, which we
have chosen such that the solution is sufficiently smooth but still
fits the kinematical data well.

\subsection{Monte Carlo Tests}

We have tested our method with artificial data, as follows: We choose
a \df\ $f^{MC}(E,L_z,K)$, constructed by fitting the deprojected
density and kinematics of NGC 1600 with a different set of basis
functions than used in the normal fitting procedure, so as to test
the ability of our basis to reproduce general distribution
functions. We calculate the projected kinematics of $f^{MC}(E,L_z,K)$,
and then draw artificial data points from the model kinematics at the
positions of the observed data points and with their respective
errors. We save $f^{MC}$ on a grid in $E,L_z$ and $K$ for later
comparision with the \df\ inferred from the artificial data.

Using the scheme described above we then obtain a solution for a 
\df, now fitting the density and the artificial kinematic
data with our normal basis.  A $\chi^2$ measure of the
deviation in the \df\ is
\[
\chi^2 = {1\over{ N_g}} \sum_{i=1}^{N_g} \left(f_i-f_i^{MC}\over f_i^{MC}\right)^2,
\]
where $N_g$ is the number of points of the grid in integral space,
$f_i$ denotes the inferred \df\ and
$f_i^{MC}$ the Monte Carlo \df\ on the $i^{th}$ grid point.

We have performed two types of Monte-Carlo tests. In the first we
place the artificial data points exactly on the predicted kinematic
profiles of the Monte Carlo model, but use the errorbars from the
observations. In this case the RMS deviation of the recovered
distribution function from the underlying model \df\ is $19\%$. The
corresponding deviation in the isotropically averaged \df\ $\bar f$ is
only $10\%$.  This is a favourable case without sampling errors.

\begin{figure*}
\dopic{\includegraphics*[width=0.9\textwidth]{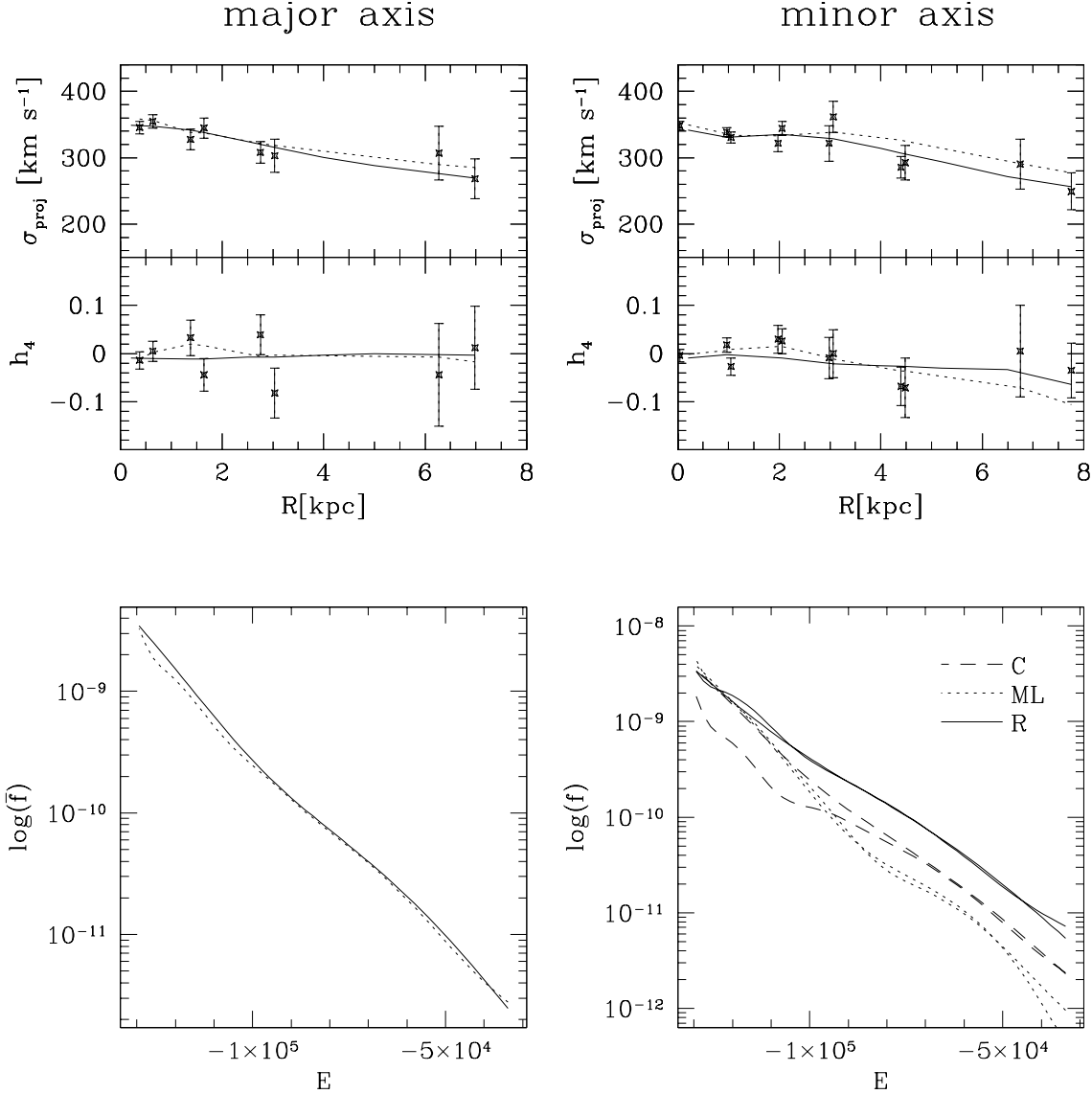}}
\caption{
\label{MCpic}
Applying the method to one of the Monte Carlo datasets.  First row:
Projected kinematics for major (left) and minor axis (right). The
solid lines represent the input model, the dotted lines show the
kinematics of the model recovered from the artificial data (points
with error bars).  Second row: In the left panel, the averaged
distribution function $\bar f(E)$ is shown for input model (solid
line) and recovered model (dotted line), as a function of energy. The
right panel shows the values of the \df\ for three typical orbits near
the radial, meridional loop, and circular orbits, again as functions
of energy.  Each line type corresponds to one of these orbits, showing
both the input and the recovered distribution function. In this
example the RMS errors of $\bar f$ and $f$ are 13\% and 27\%,
respectively.}
\end{figure*}

In the second, more realistic test the artificial data are drawn from
the Monte Carlo model as Gaussian variates with the appropiate
observational errors.  Figure~\ref{MCpic} shows, for a typical set of
Monte Carlo data, both the original \df\ (dots) and the recovered \df\
(solid lines) in this case.  Again the \df\ is recovered with good
accuracy in the entire energy range.  Typical RMS errors are $13\%$
and $27\%$ for $\bar f$ and $f$, respectively, as determined from 10
different Monte-Carlo samples.  Note that at the very centre the \df\ is
only constrained by the intrinsic density, as the kinematic data trace
the \df\ only outside a minimal radius. Therefore it is not possible
to determine the anisotropy in the very center. and this point has
been excluded in the computation of the quoted $\chi^2$ values.

\section{The dynamics of NGC 1600}

\begin{figure*}
\dopic{\center{\includegraphics[width=.76\textwidth]{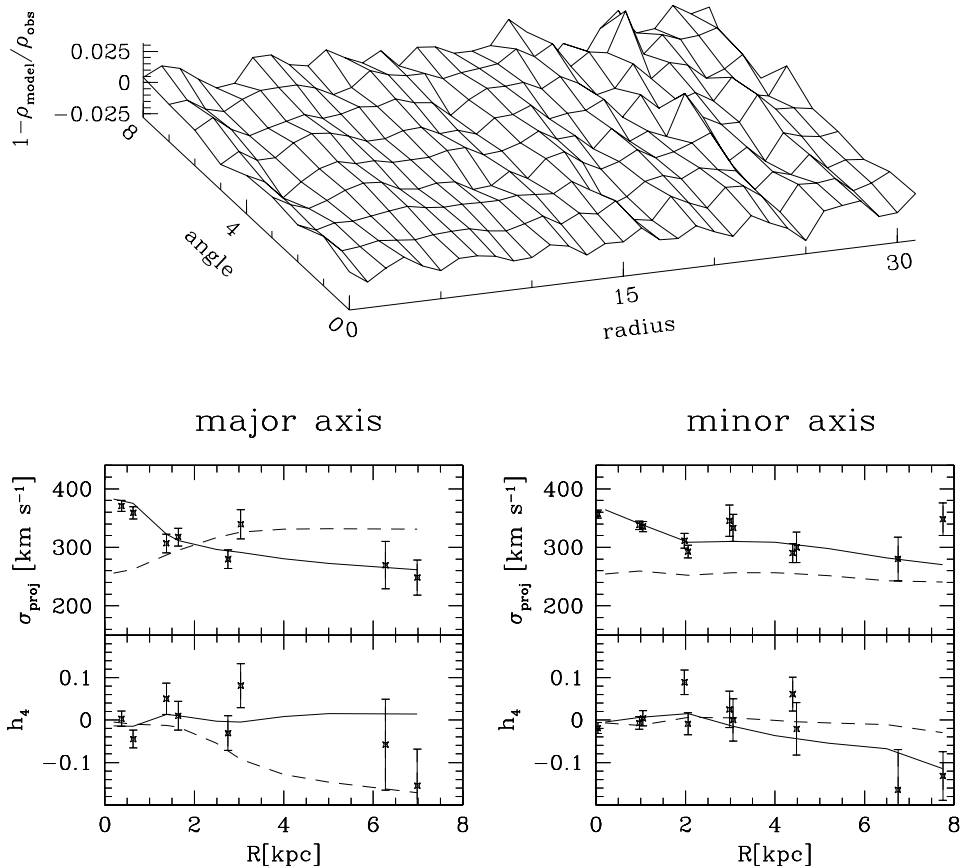}}}
\caption{\label{kinpic}
Best-fitting dynamical model for NGC 1600, derived from the deprojected
density and the line profile data of Bender \etal\ (1994).  Top:
relative deviation between deprojected and model density. Coordinates
are indices of a grid in the meridional plane. The RMS relative
deviation of the density is $1.1\%$.  Bottom: velocity dispersion and
$h_4$ parameter on the major and minor axes. The best--fit
three--integral model is shown by thick lines; for comparison, a
two--integral model $f(E,L_z)$ is also shown (dashed lines). }
\end{figure*}

\begin{figure*}
\dopic{\center{\includegraphics*[width=.9\textwidth]{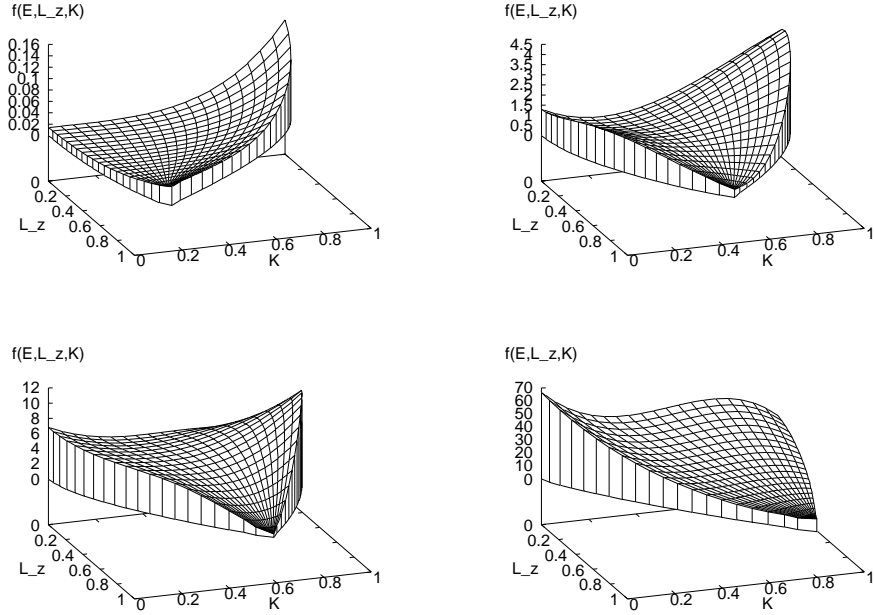}}}
\caption{\label{dfpic}
The inferred \df\ of NGC 1600 at four
different energies, parametrising shells from far out (top left)
to near the centre (bottom right).
The circular orbit radii corresponding to these energies are
$R_c=16\kpc, 2.9\kpc, 1.6\kpc, 0.5\kpc$.
Throughout most of the galaxy, the \df\ is strongly peaked on the equatorial
radial orbit (top right corner in each panel). Inside the core region
($R_b=1.85\kpc$ on the major axis) 
the model becomes less radially anisotropic.
}
\end{figure*}

\begin{figure*}
\dopic{\center{\includegraphics[width=.76\textwidth]{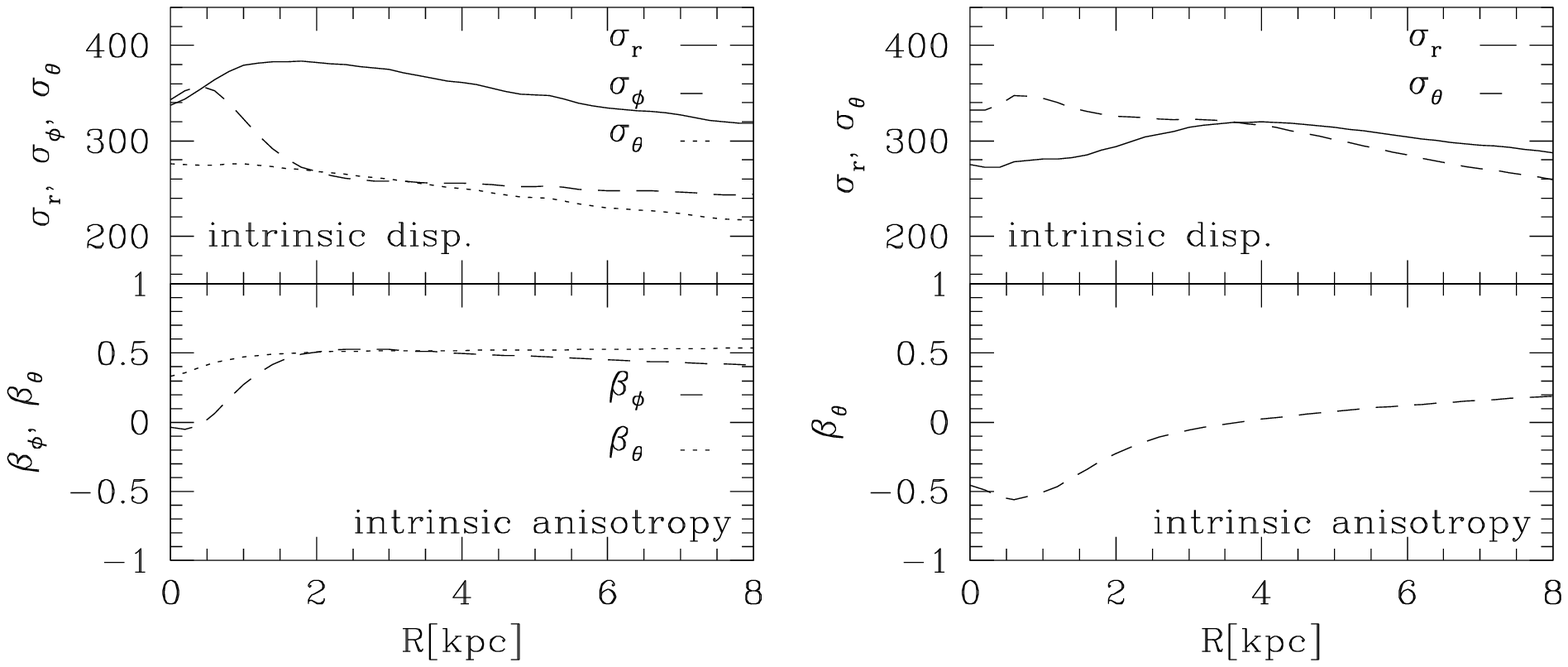}}}
\caption{\label{fitpic}
Intrinsic velocity dispersions and anisotropy parameters for NGC~1600.
Left: on $R$--axis. Right: on true minor ($z$--) axis.  The anisotropy
parameter $\beta = 1 - \sigma_t^2/\sigma_r^2$ is positive for radial
anisotropy.  }
\end{figure*}

Figures \ref{kinpic}-\ref{fitpic} show the results of applying our
method to the boxy elliptical galaxy NGC 1600. The program was asked
to fit the edge--on--deprojected density distribution, and the minor
and major axis velocity dispersions and line profile shape parameters
$h_4$ from Bender \etal\ (1994).  Fig.~\ref{kinpic} shows these data
and the best--fitting three--integral model, as well as, for
comparison, the best--fitting two--integral model.  It is clear that
NGC~1600 requires a three--integral \df. The two--integral model is a
very poor fit to the kinematic data, consistent with previous similar
but weaker results based on only velocity dispersions (van der Marel
1991, 1998).  For the three--integral model, the \rms\ relative
deviation of the density is $\approx 1.1\%$, and the fit of the
kinematic data is within one standard deviation in the mean.  There
are still slight systematic differences between our model of NGC 1600
and the kinematic data on the minor axis. If these are confirmed with
higher signal--to--noise data, this might require an inclination angle
less than $90^\circ$ or possibly a slightly triaxial potential.

These plots involve scaling the model \df\ to the data; the scaling
constant gives the mass-to-light ratio. This was found by finding the
best--fitting three--integral model for a range of values of the scaling
constant, and then determining the optimal value: we thus obtain a
best $M/L_V = 6$ and the model shown in Fig.~\ref{kinpic}.
Even for models that do not fit the kinematic data well we have always
found $M/L_V $ in the range $5.5-6.8$.

A sample of the implied phase space distribution is given in
Fig.~\ref{dfpic}. Each panel show a cut through phase space at a
fixed energy, with corresponding circular orbit radius given in the
caption.  On each energy surface $f$ is given as a function of the
angular momentum scaled to the maximal value possible at that energy
and the third integral $K$ similarly scaled.  The vertical surfaces
limit the part of integral space which is accessible to stars at this
energy. Each corner of this triangular structure represents a special
orbit: e.g., the circular orbit with the highest angular momentum is
located towards the front of the surface, and the radial orbit in the
equatorial plane is located at the top right corner.  The left
boundary represents the shell orbits, the boundary to the
right the equatorial orbits. See also Section~\ref{secISpace}.

From Fig.~\ref{dfpic} one sees that in the outer parts of the galaxy
(upper row) the radial orbits dominate. By contrast, the central
regions of the galaxy (lower row) are more isotropic, although some
radial anisotropy is still present. In the very centre (lower right
panel for circular orbit radius of $\approx 0.5 \kpc$),
meridional loop orbits are seen to dominate over equatorial radial orbits.
The transition, between radii of $0.5-1.5\kpc$, coincides with the
rise of the velocity dispersion that occurs about in this range of radii
on both axes.

To reaffirm this conclusion Fig.~\ref{fitpic} shows the inferred
intrinsic velocity dispersions and anisotropy parameters on the true
major and minor axes of NGC 1600. On the major axis the radial
dispersion $\sigma_r$ exceeds the azimuthal dispersion $\sigma_\phi$
and the meridional dispersion $\sigma_\theta$; outside the central
$\sim1.5\kpc$ the values of the two anisotropy parameters
$\beta_\phi\equiv 1-\sigma_\phi^2/\sigma_r^2$ and $\beta_\theta\equiv
1-\sigma_\theta^2/\sigma_r^2$ are $\sim 0.4$ and the model is thus
radially anisotropic, but approximately isotropic in the
$(\theta,\phi)$--plane.  At small radii along the major axis,
$\sigma_\phi$ increases to $\sim \sigma_r$, while $\sigma_\theta$
remains low.

On the minor axis, $\sigma_r$ exceeds $\sigma_\theta$ outside $z\simeq
3\kpc$. The inferred radial anisotropy is distinctly less than on the
major axis.  Moreover, at small radii along the minor axis, the
dynamical structure is reversed: there we have $\sigma_\theta >
\sigma_r$, with $\beta_\theta$ reaching $-0.5$ near the centre.  The
transition between the two regimes corresponds to the change from
predominantly equatorial radial orbits and meridional butterfly orbits
to predominantly meridional loop orbits, that occurs at circular orbit radii
around $\sim 1\kpc$ (see Fig.~\ref{dfpic}). For comparison, the radius
marking the edge of the central core region of NGC 1600 is $\sim
1.85\kpc$ on the major and $\sim 1.3\kpc$ on the minor axis.

In their study of the dynamics of three--integral oblate galaxy
models, Dehnen \& Gerhard (1993a) identified several ways of
constructing a self--consistent, flattened system. Comparing with
their results, it appears that the dynamics of NGC 1600 is closest to
their models 8 and 9, in which the flattening is achieved by putting
extra mass on equatorial radial orbits. This leads to the required
excess in the $x$- and $y$-kinetic energies compared to the kinetic
energy in the $z$-direction, and to a stronger radial anisotropy on the
major axis than on the minor axis. This pattern is similar to that
inferred above for NGC 1600, although the effect is more pronounced in
Dehnen \& Gerhard's quoted models (their model 8 is isotropic on
the minor axis). Compare Figs.~12 (\df ), 17 (velocity ellipsoids) and
18-19 (kinematics) of Dehnen \& Gerhard (1993a) and also Fig.~4
of Dehnen \& Gerhard (1993b). One characteristic for
this orbit structure is that the ratio of the measured velocity 
dispersions on the minor and major axes is significantly above
unity; for NGC 1600, $\sigma_{\rm minor}/\sigma_{\rm major}\simeq
1.15$ at $R=4\kpc$.

\section{Conclusions and Discussion}

The main result of this study is that the dynamics of NGC~1600 is
consistent with a radially anisotropic, axisymmetric three-integral
\df, in which the flattening is achieved by putting extra mass on
equatorial radial orbits. Two--integral \df s cannot reproduce the
kinematic data. The radial anisotropy is strongest in the outer parts
of the modelled range (out to $R_e/2$), with $\sigma_\theta/\sigma_r
\approx\sigma_\phi/\sigma_r\approx 0.7$ on the major axis. On the
minor axis and near the centre the galaxy is more isotropic.  The
inferred mass--to--light ratio is $M/L_V = 6 h_{50}$ with an
uncertainty of $\approx \pm 0.5 h_{50}$.

Comparing similar ground--based data with the predictions of
two--integral models, Magorrian \etal\ (1998) inferred a central
massive object of $M_\bullet\simeq 10^{10}\msun$ in NGC 1600.  Our
analysis here shows not only that a two--integral model is
inconsistent with the measured line profile shape parameters, but also
that the rise of the central velocity dispersion seen in these
ground--based data is fit well with a radially anisotropic
three--integral model without black hole (note, however, that because
of the limited radial range of the line-profile data we have used, the
contribution of high--energy radial orbits to the central velocity
dispersion peak might be overestimated). Given the overall radial
anisotropy of NGC 1600, it is likely that the black hole in NGC 1600
has a smaller mass than inferred by Magorrian \etal\ (1998). The
kinematic data used here do not discriminate for or against such
smaller black hole masses; to reduce the ambiguity in the central
gravitational potential requires high--resolution data (e.g., Kormendy
\& Richstone 1995) with small error bars.  Using HST data for NGC 3379
and orbit distribution modelling, Gebhardt \etal\ (1999) indeed found
that the implied black hole mass in their best--fitting model for NGC
3379 is about a factor of 6-7 lower than that inferred by Magorrian
\etal\ (1998).

Besides NGC 1600, radial anisotropy has also been inferred in several
E0 galaxies for which a line profile analysis has been done: NGC 2434
(Rix \etal 1997), NGC 6703 (Gerhard \etal\ 1998), NGC 1399 (Saglia
\etal\ 1999), and NGC 3379 (Gebhardt \etal\ 1999).  The model of
Dejonghe \etal\ (1996) for the flattened elliptical galaxy NGC 4697
has $\sigma_\phi>\sigma_R>\sigma_z$, whereas NGC 1700 appears to be
tangentially anisotropic (Statler \etal\ 1999); for both of these
objects no line profile data were used, however. Two S0 galaxies (NGC
3115, Emsellem \etal\ 1999, NGC 4342, Cretton \& van den Bosch 1999)
have dominant $\sigma_\phi$ dispersion.  Although the number of
galaxies investigated in enough detail is still small, there are the
beginnings of a trend in that large ellipticals appear to show a
transition from a nearly isotropic central region to a moderately
radially anisotropic main body. We are currently applying our
technique to several other elliptical galaxies to see whether this
trend holds up.

Our method is an adaptation of the techniques of Dehnen \& Gerhard
(1993a) to axisymmetric galaxy potentials. First, the galaxy is
deprojected, and the luminosity density and corresponding potential
are determined. From the potential, we calculate an approximate third
integral $K$ from perturbation theory (see Gerhard \& Saha 1991). We
then write the distribution function as a sum over basis functions in
the three integrals of motion ($E$, $L_z$, $K$), and fit to the
observed velocity moments and line profile parameters, using a
regularised non-parametric technique. The method has been validated by
recovering a model distribution function from its ``observable''
kinematics. One advantageous feature of our technique is that it directly
yields the phase--space distribution function of the galaxy. Its main
restriction is to potentials in which the third integral gives a good
approximation to the orbital tori and, if stochastic regions are
present, to their boundaries. In this way, stochastic orbit building
blocks can be constructed. In the deprojected axisymmetric potential
of NGC 1600 stochastic orbits are unimportant and resonant orbit
families take up only a minor fraction of phase space; for this galaxy
the perturbation integral gives an excellent fit to most orbits
(Fig.~\ref{sospic}).

We will now discuss our results in the light of a possible merger
origin for elliptical galaxies.  Because of the very large number of
relevant parameters no specific large-$N$ merger remnant model to
match a particular galaxy will be available for some time.  By
comparing in qualitative terms the main aspects of the dynamical
structure inferred for NGC 1600 with the dynamics of those merger
remnants that have been analyzed, we may nonetheless gain some insight
into the kind of merger process that may have shaped this galaxy.

The dynamical properties of the remnants in the published merger
calculations depend strongly on the assumed physics and initial
conditions. Mergers between two purely stellar, about equal-mass disk
galaxies typically result in remnants with large triaxiality and
kinematic misalignment (Barnes 1992, Heyl, Hernquist \& Spergel 1996),
unless the encounter is a prograde one with relatively large impact
parameter. Remnants of minor mergers rotate significantly (Barnes 1998).
Including a gaseous component in the simulations, even
with only a small fraction of the total mass, results in significantly
more oblate remnants and in a smaller fraction of box orbits relative
to tube orbits; however, the difference in orbital structure due to
different numerical algorithms is substantial (Fig.~17 of Barnes \&
Hernquist 1996).

Another possible process is the merging of several smaller
parts to one large galaxy (Weil \& Hernquist 1996, Dubinski 1998).
Again, the initial trajectories of the merging galaxies have an
important impact on the remnant. The remnants resulting from nearly
isotropic initial conditions are clearly more axisymmetric and rounder
than pair merger remnants, but show large rotation velocities (Weil
\& Hernquist 1996).  Dubinski (1998) used CDM simulations to
get initial conditions at $z=2$.  From that time on, he followed the
merging process to a brightest cluster galaxy. In his simulation,
merging pieces fall in mainly along filaments. The resulting remnant
has a triaxial shape aligned with its environment, it shows only slow
rotation around the small axis, and is mildly radially anisotropic
with anisotropy parameter increasing slowly outwards.

The lack of rotation in NGC 1600 clearly argues against a binary
merger on a wide orbit or a merger of a near-isotropic galaxy group as
simulated by Weil \& Hernquist (1996).  We do not know how strongly
triaxial NGC 1600 is (the argument given in the Introduction is
statistical, and the kinematic misalignment in this galaxy is not
well-defined). However, our derived \df\ for NGC 1600
(Fig.~\ref{dfpic}) has a strong bias towards radial (z-tube)
orbits. By contrast, the angular momentum distributions for the {\sl
z-tube orbits alone} in the merger remnants analysed by Barnes (1992)
and Barnes \& Hernquist (1996) are fairly uniform. The inferred large
extra mass on radial orbits in NGC 1600 may then either correspond to
the large fraction of box orbits present in some of these remnants (if
NGC 1600 is significantly triaxial), or it may have evolved out of
such box orbits if the shape of NGC 1600 has evolved towards
axisymmetry since its formation. In either case, the large radial
orbit fraction argues for a merger where the effects of gas were not
very important. This argument is also supported by the large observed
core radius of NGC 1600.  Thus the dynamics of NGC 1600 appear
consistent both with a mainly collisionless, low-angular momentum
binary merger and with a variant of the merging--along--filaments
described by Dubinski (1998).

It will be interesting to address such questions for several more
elliptical galaxies, hopefully with a larger sample of quantitatively
analysed merger remnants at hand.  In our view, for these comparisons
the most helpful structural information about the merger remnants
will be their three-dimensional velocity ellipsoids.

\section*{Acknowledgements}

We thank Ralf Bender for sending us his unpublished surface
photometry, Walter Dehnen for the use of his programs, David Merritt
for a discussion about regularisation techniques, and the referee, Tim
de Zeeuw, for comments that improved the presentation of this
paper. We gratefully acknowledge support by the Swiss National Fund
under grants 20-43218.95 and 20-50676.97.




\label{lastpage}

\end{document}